\documentstyle[prd,aps,epsf]{revtex}
\begin{document}

\twocolumn[\hsize\textwidth\columnwidth\hsize\csname
@twocolumnfalse\endcsname

\title{Anti-de Sitter no-hair, AdS/CFT and the brane-world}

\author{Tetsuya Shiromizu}
\address{
Research Centre for the Early Universe(RESCEU), \\
The University of Tokyo, Tokyo 113-0033, Japan}

\author{Daisuke Ida}
\address{Department of Physics, Kyoto University, Kyoto 606-8502, Japan}


\date{\today}

\maketitle

\begin{abstract}
We study the asymptotic behavior of the bulk spacetimes with the
negative cosmological constant in the context of the brane-world scenario.
We show that, in Euclidean bulk, or in Lorentzian static bulk,
some sequences of hypersurfaces with the positive Ricci scalar evolve to 
the warped geometries like the anti-de Sitter spacetime. 
Based on the AdS/CFT correspondence,
we discuss that the positivity of the Ricci scalar is related to
the stability of CFT on the brane. In addition, the brane-world is 
described from the holographic point of view. The asymptotic local 
structure of the conformal infinity is also investigated. 
\end{abstract}
\vskip2pc]

\vskip1cm

\section{Introduction}

The recent progress of M/String theory suggests 
 a new picture
that our universe is described by a domain wall in a higher
dimensional spacetime.
The simplest and comprehensive model have
been proposed by Randall and Sundrum \cite{RS1,RS2}. Therein our
universe is realised as the three-brane (four-dimensional
Minkowski spacetime) embedded
in the five-dimensional anti-de Sitter (AdS) bulk spacetime.

The brane has the fine-tuned non-zero tension to
cancel the bulk cosmological constant. One might
expect that the brane inherits the gravity from the
higher-dimensional bulk spacetime. We can, however,
check that the gravity is
indeed localised on the brane;
namely the usual Newton law of
gravitation is realised at least on the low energy scales
\cite{RS1,RS2,Tama,Sasaki,Tess,Lisa}. There is also the
exact cosmological solution based on the brane-world scenario,
which is given by the domain wall motion in the five-dimensional
Schwarzshild--AdS
spacetime \cite{cosmos,Misao}.
The warped compactification is also used
in the holographic renormalisation\cite{Holo}.
In the same sense as Witten suggested,
the brane-world contains the aspect of the AdS/CFT
correspondence \cite{AdSCFT},
in which the bulk graviton
is identical to CFT with the cut-off on the brane 
\cite{Lisa,AdS1,AdS2,AdS3}.

In this brane-world scenario, the
warped geometry is essential for the
mass hierarchy problem \cite{RS1}
and the localisation of gravity \cite{RS2}.
This is also preferred for the cosmological constant problem\cite{Izawa}.
This mechanism can be contrasted with that of the inflationary scenario,
in which
the rapid cosmological expansion in the early universe
resolves the homogeneity, horizon and flatness problems\cite{inflation}.
It is believed that an initially inhomogeneous universe evolves to a
homogeneous isotropic state during the inflation.
In other words, there is the so-called cosmic no-hair (baldness) conjecture,
which states that all expanding universes approach at least locally
the de Sitter spacetime, if there is a positive cosmological constant.
The cosmic no-hair conjecture is confirmed in several cases
\cite{boucher,Wald}.

Bearing the inflation in mind,
one might expect that the warped geometry of the brane-world
model is the general effect of the negative cosmological constant;
namely, the bulk spacetime might be governed by
{\em the AdS no-hair theorem} in some sense,
and this is the subject we consider here.
We will adopt the following geometrical approach;
the brane-world is 
just the slicing of hypersurfaces in asymptotically 
AdS spacetimes. The slice or several slices are regarded as
the branes where we or hidden matters are living. The extrinsic
curvature determines the matter distribution including the tension
on the brane. In this geometrical picture,
it is important to study
the general features of asymptotic structure.

This paper is organised as follows. In section \ref{II} we
show that the bulk geometry foliated by probe branes with
positive or zero Ricci curvature approaches the standard
AdS form asymptotically in the cases of the static bulk and
the Euclidean bulk.
Then, we consider the validity of the assumption $R \geq 0$,
and discuss the relation to the AdS/CFT correspondence.
In section \ref{III}, we also show the
asymptotic local structure of the conformal infinity.
The features obtained there are appropriate
for the mass hierarchy problem and so on.

\section{No-Hair and AdS/CFT}\label{II}

Our strategy is as follows: First, we consider the full geometry
regardless of the branes. Then, choosing slices appropriately,
we may be able to identify them with the branes where we or hidden
matters are living. After some cutting and pasting, 
this results in the brane-world model.

\subsection{The No-Hair}

First of all, we show that the bulk geometry ``evolve'' to
the AdS spacetime.  Here we consider the ``evolution'' of the bulk in the
Euclidean case\footnote{For example, the holographic renormalisation group
and the AdS/CFT correspondence are
often formulated in the Euclidean spacetime which offers the
regular boundary condition.
In this sense, the consideration of
the Euclidean bulk is significant.
In general, we may expect that the consideration of the Euclidean region
gives us the relevant boundary condition\cite{Euclid}.}
and the Lorentzian static case. The
discussion in this subsection is reminiscent
of the Wald's cosmic no-hair theorem for
homogeneous universe with negative Ricci spatial curvature
(namely, except for the Bianchi IX model) \cite{Wald}.

We assume that the bulk is foliated by the geodesic slices\footnote{We know
that this foliation can work very well locally.
But, in general cases, we cannot use it globally because
the focusing point occurs due to the presence of inhomogeneities.}:
%
\begin{eqnarray}
{}^{(5)}g_{ab}dx^adx^b=dy^2+g_{\mu\nu}(y,x)dx^\mu dx^\nu,
\end{eqnarray}
%
where $\{x^{\mu}\}=\{x^0,\cdots,x^3\}$ and $y=x^4$ is the coordinate of 
the extra dimension. We have two basic equations: One
is the equation for the trace of the extrinsic curvature, $K$;
%
\begin{eqnarray}
\dot K =\frac{4}{\ell^2}-\frac{1}{4}K^2-\sigma_{\mu\nu}\sigma^{\mu\nu},
\label{eq:Ray}
\end{eqnarray}
%
where the dot denotes the derivative with respect to $y$, and
\begin{equation}
\sigma_{\mu\nu}=K_{\mu\nu}-{\textstyle \frac{1}{4}}g_{\mu\nu}K
\end{equation}
is the trace-free part of the extrinsic curvature. Another is
the Hamiltonian constraint:
%
\begin{eqnarray}
R-\frac{3}{4}K^2+\sigma_{\mu\nu}\sigma^{\mu\nu}=-\frac{12}{\ell^2}.
\label{eq:Hami}
\end{eqnarray}
%
In the above derivations, we have assumed the five-dimensional
Einstein equation for the bulk:
%
\begin{eqnarray}
{}^{(5)}R_{ab}-{\textstyle
\frac{1}{2}}{}^{(5)}g_{ab}{}^{(5)}R=-\frac{6}{\ell^2}{}^{(5)}g_{ab}.
\end{eqnarray}
%
In the Euclidean, and the Lorentzian static cases\footnote{In the static
case,
$\sigma_{0i}=0$ for $i=1,2,3$.}, we can see that the inequality
%
\begin{eqnarray}
\sigma_{\mu\nu}\sigma^{\mu\nu}\geq 0
\end{eqnarray}
%
holds.
Using Eq~(\ref{eq:Ray}) together with the above inequality
and assuming
$R\geq 0$ and $K>0$ ($y=y_0$ chosen arbitrarily), we obtain 
%
\begin{eqnarray}
K \to \frac{4}{\ell}.~~~(y\to+\infty) \label{eq:trace}
\end{eqnarray}
%
The Hamiltonian constraint (\ref{eq:Hami}) therefore implies
%
\begin{eqnarray}
\sigma_{\mu\nu}\to 0.~~~(y\to+\infty)
\label{eq:shear}
\end{eqnarray}
%
{}From Eqs~(\ref{eq:trace}) and (\ref{eq:shear}), the metric of
slices behaves as
%
\begin{eqnarray}
g_{\mu\nu}(y,x) \to e^{(y-y_0)/\ell}h_{\mu\nu}(y_0, x).~~~(y\to+\infty)
\end{eqnarray}
%
Then we naively conclude that the bulk geometry evolve toward the warped 
geometries like the AdS spacetime. Since the metric of the terminal
hypersurface $\{y=y_0\}$ is not fixed in the above argument, the bulk
geometry does not necessarily evolve to
the exact AdS.

This theorem can be easily extended to the cases with bulk fields:
Eqs (\ref{eq:Ray}) and (\ref{eq:Hami}) are modified as
%
\begin{eqnarray}
\dot K & = & \frac{4}{\ell^2}-\frac{1}{4}K^2-\sigma_{\mu\nu}\sigma^{\mu\nu}
\nonumber \\
& & ~~-\kappa_5^2\Bigl(T_{ab}^{({\rm bulk})}-{\textstyle
\frac{1}{3}}{}^{(5)}g_{ab}
T^{({\rm bulk})}  \Bigr)n^a n^b ,
\end{eqnarray}
%
and
%
\begin{eqnarray}
R-\frac{3}{4}K^2+\sigma_{\mu\nu}\sigma^{\mu\nu}
+2\kappa_5^2 T_{ab}^{({\rm bulk})}n^a n^b =-\frac{12}{\ell^2},
\end{eqnarray}
%
where $n=\partial_y$, and we have used
%
\begin{eqnarray}
{}^{(5)}R_{ab}-{\textstyle \frac{1}{2}}{}^{(5)}g_{ab}{}^{(5)}R
=-\frac{6}{\ell^2}g_{ab}
+\kappa_5^2T_{ab}^{({\rm bulk})}.
\end{eqnarray}
%
If we suppose the additional conditions
%
\begin{eqnarray}
T_{ab}^{({\rm bulk})}n^a n^b \geq 0 \label{eq:energy}
\end{eqnarray}
%
and
%
\begin{eqnarray}
\Bigl(T_{ab}^{({\rm bulk})}-{\textstyle \frac{1}{3}}{}^{(5)}g_{ab}
T^{({\rm bulk})}  \Bigr)n^a n^b \geq 0,\label{eq:energy2}
\end{eqnarray}
%
then we obtain the similar theorem. The condition Eq (\ref{eq:energy}) is 
naively identical with the reasonable condition that the effective 
pressure is positive in the bulk. 
On the other hand, the condition Eq (\ref{eq:energy2}) requires 
the condition $\rho-P \geq 0$ for the effective pressure $P$ and 
energy density $\rho$ in the bulk.

We have assumed that
$ R \geq 0$ holds to show the AdS no-hair theorem. In the
next subsection, we examine this assumption in several aspects.

Finally we describes two examples for no-hair argument. 
(i) five-dimensional Schwarzshild-AdS spacetime: The metric is
%
\begin{eqnarray}
ds^2_5=-f(r)dt^2+f^{-1}(r)dr^2+r^2 d\Sigma_k^2,
\end{eqnarray}
%
where $f(r)=k-\mu/r^2 +(r/\ell)^2$ and $k=0,\pm 1$.
Let us consider the foliation
of $r={\rm constant}$ hypersurfaces having the Ricci scalar:
%
\begin{eqnarray}
R=\frac{6k}{r^2}.
\end{eqnarray}
%
The trace of extrinsic curvature is
%
\begin{eqnarray}
K=\frac{4}{\ell}\Bigl[1+\Bigl( \frac{\ell}{r}\Bigr)^2+\cdots  \Bigr].
\end{eqnarray}
%
The shear becomes
%
\begin{eqnarray}
\sigma_{tt}=-\frac{3k}{4\ell}\Bigl( \frac{\ell}{r}\Bigr)^2
\end{eqnarray}
%
and
%
\begin{eqnarray}
\sigma_{ij}=\frac{k}{4\ell}\Bigl( \frac{\ell}{r}\Bigr)^2 \delta_{ij}.
\end{eqnarray}
%
Regardless of the signature of
$R$, the geometry evolve to the AdS-like geometry.

(ii)Sol Bianchi ${\rm VI}_{-1}$: The metric is\cite{solution}
%
\begin{eqnarray}
ds^2_5& = & -\frac{4}{3}\frac{r^2}{\ell^2}dt^2
+\frac{3}{4}\frac{\ell^2}{r^2}dr^2 \nonumber \\
& & ~~+\frac{1}{2}\ell^2[r^2(e^{2z}dx^2+e^{-2z}dy^2 )+dz^2].
\end{eqnarray}
%
Let us consider the foliation of $r={\rm constant}$ hypersurfaces having
the Ricci scalar:
%
\begin{eqnarray}
R=-\frac{4}{\ell^2}.
\end{eqnarray}
%
The trace of extrinsic curvature is
%
\begin{eqnarray}
K=2{\sqrt {3}} \frac{1}{\ell}.
\end{eqnarray}
%
The Ricci scalar is negative and $K$ does not approach to the value
$4/\ell$. This is
a case which will not evolve to AdS-like geometry.

\subsection{AdS/CFT and the Ricci Curvature on the Brane}

To see the signature of the Ricci scalar,
$R$, on each
hypersurfaces, we use the gravitational equation on the brane\cite{Tess}.
It is derived using the Gauss equation, which connects the five-dimensional
Riemann tensor
with that in four dimensions, and the Israel's junction
condition\cite{Israel}:
%
\begin{eqnarray}
G_{\mu\nu}=8\pi G_4 T_{\mu\nu}+\kappa_5^4 \pi_{\mu\nu}-E_{\mu\nu},
\end{eqnarray}
%
where
%
\begin{eqnarray}
\pi_{\mu\nu} & = &
-\frac{1}{4}T_{\mu\alpha}T^\alpha_\nu+\frac{1}{12}TT_{\mu\nu} \nonumber \\
& & ~~+\frac{1}{8}g_{\mu\nu}T_{\alpha\beta}T^{\alpha\beta}
-\frac{1}{24}g_{\mu\nu}T^2
\end{eqnarray}
%
and
%
\begin{eqnarray}
E_{\mu\nu}={}^{(5)}C_{\mu\alpha\nu\beta} n^\alpha n^\beta.
\end{eqnarray}
%
We set the net cosmological constant on the brane to be zero.
To evaluate $E_{\mu\nu}$ we have to solve it globally\cite{Sasaki,Tess,Gen}.
The trace part can be written in terms of the four-dimensional
quantities:
%
\begin{eqnarray}
R=-8\pi G_4 T-\frac{\kappa_5^4}{4}\Bigl(T_{\mu\nu}T^{\mu\nu}
-\frac{1}{3}T^2 \Bigr)\label{eq:effective}
\end{eqnarray}
%
On the other hand, from AdS/CFT, we can obtain the following effective
equation on the brane\cite{Lisa,AdS1,AdS2,AdS3,Ano,Nojiri,Olsen}:
%
\begin{eqnarray}
G^{\mu\nu}=8\pi G_4 T^{\mu\nu}+\frac{4}{\ell}\frac{1}{{\sqrt {|g|}}}
\Bigl( \frac{\delta S^{(4)}
_{\rm ct}}{\delta g_{\mu\nu}} +
\frac{\delta \Gamma_{\rm CFT}}{\delta g_{\mu\nu}}  \Bigr),
\label{19}
\end{eqnarray}
%
where $\Gamma_{\rm CFT}$ is the effective action of CFT living on the
boundary and has the trace anomaly\cite{AdS2,AdS3,Ano}:
%
\begin{eqnarray}
\frac{\delta \Gamma_{\rm CFT}}{\delta g_{\mu\nu}}g^{\mu\nu}=
\frac{\ell^3}{16}
{\sqrt {|g|}}\Bigl(
R_{\mu\nu}
R^{\mu\nu}-{\textstyle \frac{1}{3}}
R^2  \Bigr). \label{eq:anomaly}
\end{eqnarray}
%
The quantity
$S_{\rm ct}^{(4)}$
 is $R^2$ terms of the counter-term which makes the action finite. For
our purpose, we need not to write down the explicit form
(See Ref. \cite{AdS3}.).
What we will use is that $\delta S_{\rm ct}/\delta g_{\mu\nu}$ is traceless.
%
%
Then, the trace part of Eq (\ref{19}) is
%
\begin{eqnarray}
R=-8\pi G_4 T-\frac{\ell^2}{4}
\Bigl(
R_{\mu\nu}
R^{\mu\nu}-{\textstyle \frac{1} {3}}
R^2  \Bigr)
\label{eq:CFT}
\end{eqnarray}
%
Using the Einstein equation,
$G_{\mu\nu}=8\pi G_4 T_{\mu\nu}$,
approximately, we can check that Eq (\ref{eq:effective}) is 
approximately identical with
Eq (\ref{eq:CFT}). It is remarkable that $\rho^2$ terms appeared in
the cosmological solution of the brane-world can be regarded as
the non-linear contribution from CFT\footnote{Some class of
the homogeneous-isotropic
cosmological solution is described by the motion of the domain wall in the
five-dimensional AdS spacetime. In this case, $E_{\mu\nu}=0$ and
the equation for the scale factor can be derived from
Eq (\ref{eq:effective}).}. In the linear order we easily obtain
the relation between the energy-momentum tensor of CFT and a part of
Weyl tensor:
%
\begin{eqnarray}
E^{\mu\nu} \simeq - \frac{4}{\ell}\frac{1}{{\sqrt {|g|}}}
\frac{\delta \Gamma_{\rm CFT}}{\delta g_{\mu\nu}}.
\end{eqnarray}
%
This has been suggested by Witten and firstly confirmed by
Gubser\cite{AdS1}.
The elegant formulation has
been done in Refs.~\cite{Lisa,AdS2,AdS3}. This is applied to the
cosmology with large $N$ CFT \cite{New}. The generalised Friedmann equation
has the dark-radiation term, which comes from the mass term of the
five-dimensional Schwarzshild-AdS spacetime. Indeed,
we can check that the
entropy of the dark radiation is exactly same as the black-hole entropy
\cite{Misao}. This is a realisation of the AdS/CFT in the brane-world.

{}From now on we consider the signature of $R$.
For simplicity, we consider the massless scalar field on the brane:
%
\begin{eqnarray}
T_{\mu\nu}=\partial_\mu \varphi \partial_\nu \varphi -{\textstyle
\frac{1}{2}}g_{\mu\nu}
\partial_\lambda \varphi \partial^\lambda \varphi
\end{eqnarray}
%
%
\begin{eqnarray}
T^\mu_\mu=-\partial^\mu \varphi \partial_\mu \varphi \leq 0
\end{eqnarray}
%
and
%
\begin{eqnarray}
\pi^\mu_\mu={\textstyle \frac{1}{6}}
(\partial_\mu \varphi \partial^\mu \varphi)^2 \geq 0.
\end{eqnarray}
%
In the linear order of $T_{\mu\nu}$ we see that $R \geq 0$ holds.
However, $\pi^\mu_\mu \sim \rho^2$ term dominates
in the very early universe. 
As stated before, $\pi^\mu_\mu$ is related to the trace anomaly of
CFT living on the brane. This means that $\pi^\mu_\mu$ is
quantum effect in some sense. To check the signature of $T^\mu_\mu$
and $\pi^\mu_\mu$, 
we must consider the concrete model of the matter on the brane.
This seems to be hard task in general.

We may be able to use the Witten-Yau argument \cite{Ricci,Models}, where
they showed that the boundary with the positive Ricci scalar will be
stable for large $y$(near the infinity). They discussed the fluctuation 
of the brane which the mass term of the fluctuation mode is proportional to 
the Ricci scalar on the brane. 
This therefore seems to justify the assumption of the
positive Ricci scalar. At first glance, however,
their argument on the stability seems inappropriate
for the
present purpose, because
they does not include the higher curvature term, $R^2$.
Fortunately, it suddenly turns out that
this is not the case for four-dimensional branes. You can check
that, in four-dimensions, the Witten-Yau argument does not depend on whether
$R^2$ term is included.

One might want to consider only background geometry because the
holographic argument is often concentrated on the probe brane
in the fixed background. In this case, we can set $T_{\mu\nu}=0$,
which automatically implies $R=0$. Thus, we can anyway conclude
that a no-hair theorem
described here holds at least for the empty brane case.

\section{The Conformal Infinity 
and The No-Hair}\label{III}

In this section, we assume the existence of the infinity and
we briefly consider the detail of the asymptotic structure
of the conformal infinity. Note that we did not assume the
existence in the previous section. To discuss the geometry near
the infinity, we will use the
so-called conformal completion formalism
\cite{deSitter1,deSitter2,deSitter3}.
We focus on the
local differential geometrical structure near the infinity
\footnote{Apart from asymptotically flat
spacetimes \cite{flat}, the spacetimes with the cosmological constant needs
the
extra assumption to derive the asymptotic symmetry and the conserved
energy \cite{deSitter2}. We will not discuss this point.}. We do 
not impose the positive Ricci scalar condition.

{\it Definition:}
\footnote{This definition is straightforward one from 
the definition of asymptotically 4-dimensional AdS spacetimes
\cite{deSitter1,deSitter2,deSitter3}. 
It should be noted, however, that this definition is rather week 
so that the black-string solutions\cite{string} are included.} 
An $n$-dimensional spacetime $(M,{}^{(n)}g)$  will be said
to be {\it
asymptotically AdS} if there exists a manifold with boundary,
$\hat M$, with the metric $\hat g$ such that: (i) there exists a function
$\Omega$ on $M$ such that $\hat g =\Omega^2 {}^{(n)}g$ on $M$; (ii) ${\cal
I}=\partial
\hat M$ and $\Omega=0$ on ${\cal I}$;  (iii) ${}^{(n)}g$ satisfies
${}^{(n)}R_{ab}-(1/2){}^{(n)}g_{ab}{}^{(n)}R+\Lambda {}^{(n)}g_{ab}=
8\pi G_n  T_{ab}$, where $\Omega^{-(n-1)}T_a^b$ admits a smooth limit to
${\cal I}$ and $\Lambda$ is the negative cosmological constant.

After some calculations, we can see
%
\begin{eqnarray}
\hat n^\mu \hat n_\mu & = & -\frac{2}{(n-1)(n-2)}\Lambda+
\frac{\Omega^2}{n(n-1)}\hat R+O(\Omega^3) \nonumber \\
& = & \hat N_0^2 +
\frac{\Omega^2}{n(n-1)}\hat R+O(\Omega^3)
\end{eqnarray}
%
and
%
\begin{eqnarray}
& &\Bigl( \hat \nabla_a \hat n_b-\frac{1}{n}{}^{(n)}g_{ab}\hat \nabla_c \hat
n^c
\Bigr)
(\hat e_{\mu})^a(\hat e_{\nu})^b
\nonumber \\
& & ~~~=
\hat \nabla_\mu \hat n_\nu -\frac{1}{n}{}^{(n)}\eta_{\mu\nu} \hat
\nabla_\alpha
\hat n^\alpha \nonumber \\
& & ~~~=  -\frac{\Omega}{n-2}\Bigl( \hat R_{\mu\nu}
-\frac{1}{n}\eta_{\mu\nu} \hat R \Bigr)+O(\Omega^2),
\end{eqnarray}
%
where $\hat n^a=\hat g^{ab}\hat \nabla_b \Omega $ and
$\{\hat e_{\mu}\}$ are the orthonormal base vectors of
$(M, \hat g)$. Then we can
obtain the asymptotic behavior of the expansion rate
$\Theta$ and shear $\Sigma_{ab}$ (see Appendix A):
%
\begin{eqnarray}
\Theta & := & \nabla_a \bar n^a \nonumber \\
& = & -(n-1)\hat N_0-\frac{n+1}{2n(n-1)\hat N_0}\Omega^2
\hat R+O(\Omega^3) \label{eq:expansion}
\end{eqnarray}
%
and
%
\begin{eqnarray}
\Sigma_{\mu\nu}& = & \Sigma_{ab}(e_{\mu})^a(e_{\nu})^b \nonumber \\
& =&  \Bigl[q_{ac}\nabla^c \bar n_b -\frac{1}{n-1}q_{ab}
\nabla_c \bar n^c\Bigr](e_{\mu})^a(e_{\nu})^b   \nonumber \\
& = & O(\Omega^2),
\label{eq:fluc}
\end{eqnarray}
%
where $\{e_{\mu}\}$ are the orthonormal base vectors of
$(M,{}^{(n)}g)$, $\bar n^a$ is the unit normal vectors defined by
%
\begin{eqnarray}
\bar n^a=\frac{{}^{(n)}g^{ab}\nabla_b
\Omega}{(\nabla_c \Omega \nabla^c \Omega )^{1/2}}
\end{eqnarray}
%
and $q_{ab}={}^{(n)}g_{ab}-\bar n_a \bar n_b$.
{}From 
the above formulas, we can conclude that the geometry near infinity has the
similar structure to that of the AdS spacetime.

Finally, we investigate on the behavior of the Weyl tensor.
To do so we use the useful equation derived from
the Bianchi identity\cite{deSitter2};
%
\begin{eqnarray}
& & \Omega \hat \nabla_{[a}\hat S_{b]c}+\Omega^{-2}\hat g_{c[a}L_{b]d}\hat
n^d
\nonumber \\
& &~~~ +\frac{n-2}{2}\hat C_{abcd} \hat n^d=\hat
\nabla_{[a}(\Omega^{-1}L_{b]c}),
\label{31}
\end{eqnarray}
%
where
%
\begin{eqnarray}
S_{ab}= {}^{(n)}\hat R_{ab}-\frac{1}{2(n-1)}\hat g_{ab}{}^{(n)}\hat R
\end{eqnarray}
%
and
%
\begin{eqnarray}
L_{ab}=\Omega^2
\Bigl( {}^{(n)}R_{ab}-\frac{1}{2(n-1)}{}^{(n)}g_{ab}{}^{(n)}R \Bigr).
\end{eqnarray}
%
{}From Eq~(\ref{31}), we obtain
%
\begin{eqnarray}
{}^{(n)}C_{\mu\nu\lambda\rho}\bar n^\rho=0,
\end{eqnarray}
%
and then
%
\begin{eqnarray}
{}^{(n)}E_{\mu\nu}&:=&{}^{(n)}C_{\mu\lambda\nu\rho}\bar n^\lambda \bar
n^\rho=0,\\
%
%
{}^{(n)}B_{\mu\nu\lambda}&:=&
q_{\mu}{}^{\rho}
{}^{(n)}C_{\rho\nu\lambda\sigma}\bar n^\sigma  =0,
\end{eqnarray}
%
on ${\cal I}$.
Note however
%
\begin{eqnarray}
{}^{(n)}C_{\mu\nu\lambda\rho}\neq 0,
\end{eqnarray}
%
in general even on ${\cal I}$.
There is the substantial difference from the
$(n=4)$-dimensions, in which ${}^{(4)}C_{\mu\nu\lambda\rho}=0$
on ${\cal I}$ \cite{deSitter2}.

\section{Summary}

In this paper, we have shown that the bulk spacetimes with negative
cosmological constant naively evolve to the anti-de Sitter
spacetime. However, the induced structure of the brane cannot
be fixed. What we can conclude is only that the local structure
on the brane will be relatively spread due to the negative
cosmological constant. This statement is useful to
discuss the mass hierarchy problem, the localisation of gravity
and the cosmological constant problem. 

In order to show the no-hair theorem, we have used the positivity
condition of the Ricci scalar of probe branes. This
might be justified via the Witten-Yau's stability argument \cite{Ricci}.
It is stressed that the no-hair theorem is valid when we 
consider the foliation by empty probe branes.

\section*{Acknowledgements}

This work was inspired by K. Izawa's talk given in the RESCEU workshop
``Brane World'' at Tokyo.
TS would like to thank Y. Imamura, K. Izawa, Y. Mino, H. Ochiai
and J. Soda for discussions. DI would like to thank H. Sato,
K. Nakamura and K. Yasuno for discussions.
TS's work is partially supported by Yamada Science Foundation.
DI was supported by JSPS Research,
and this research was supported in part by the Grant-in-Aid for Scientific
Research Fund
(No. 4318).

\appendix

\section{Calculations}

The conformal transformation of Ricci tensors gives us
%
\begin{eqnarray}
& & \Omega^{-2}\Bigl(R_{\mu\nu}
-\frac{1}{2}\eta_{\mu\nu}R+\Lambda \eta_{\mu\nu} \Bigr) \nonumber \\
& & ~~=\hat R_{\mu\nu}-\frac{1}{2}\eta_{\mu\nu}\hat R
+(n-2)\Omega^{-1}\hat \nabla_\mu \hat \nabla_\nu \Omega \nonumber \\
& & ~~-(n-2)\Omega^{-1}\eta_{\mu\nu}\hat \nabla_\alpha \hat \nabla^\alpha
\Omega \nonumber \\
& & ~~+\frac{(n-1)(n-2)}{2}\Omega^{-2}\hat \nabla_\alpha \Omega
\hat \nabla^\alpha \Omega \eta_{\mu\nu}  +\Omega^{-2}\Lambda \eta_{\mu\nu}.
\label{eq:A1}
\end{eqnarray}
%
Looking at the $\Omega$-dependence in Eq (\ref{eq:A1}), we see
%
\begin{eqnarray}
\hat N^2:=
\hat g_{\mu\nu} \hat n^\mu \hat n^\nu=-\frac{2\Lambda}{(n-1)(n-2)}+O(\Omega)
\end{eqnarray}
%
%
\begin{eqnarray}
\hat \nabla_\mu \hat n_\nu -\frac{1}{n}\eta_{\mu\nu}\hat \nabla_\alpha
\hat n^\alpha & = & \frac{1}{n-2}\Omega^{-1}
\Bigl( R_{\mu\nu}-\frac{1}{n}\eta_{\mu\nu}R \Bigr) \nonumber \\
& & -\frac{\Omega}{n-2}\Bigl(\hat R_{\mu\nu}
-\frac{1}{n}\eta_{\mu\nu}\hat R \Bigr)
\end{eqnarray}
%
and
%
\begin{eqnarray}
\hat \nabla_\mu \hat n^\mu & = & -\frac{1}{2(n-1)}\Omega \hat R
-\frac{1}{(n-1)(n-2)}\kappa_n^2T\Omega^{-1} \nonumber \\
& & ~~+\frac{n}{2}\Omega^{-1}\Bigl[\hat n_\mu \hat n^\mu
+\frac{2}{(n-1)(n-2)}\Lambda \Bigr].
\end{eqnarray}
%

Using the conformal rescaling,
$\Omega \longrightarrow \tilde \Omega = \Omega \omega$,
we obtain the following transformation:
%
\begin{eqnarray}
\tilde \nabla_a \tilde n^a & = & \frac{1}{\omega}\hat \nabla_a \hat n^a
+\frac{n}{\omega^2}\hat n^a \hat \nabla_a \omega \nonumber \\
& & +(n-2)\Omega \omega^{-3}\hat \nabla_a \omega \hat \nabla^a \omega
+\frac{\Omega}{\omega^2}\hat \nabla_a \hat \nabla^a \omega.
\end{eqnarray}
%
Let us choose $\omega$ so that $n\omega^{-1}\hat n^a \hat \nabla_a \omega
=-\hat \nabla_a \hat n^a$. Then we are resulted in
$\hat \nabla_a \hat n^a=O(\Omega)$. Moreover, using of the rescaling,
$\omega=\omega_0 +\Omega^2 f(x)$
such that $\hat n^a \hat \nabla_a \omega_0=0$,
and choosing the function $f(x)$ appropriately, we can show
%
\begin{eqnarray}
\hat \nabla_\mu \hat n^\mu =O(\Omega^2).
\end{eqnarray}
%
{}From Eq (\ref{eq:A1}) we then see
%
\begin{eqnarray}
\hat n_\alpha \hat n^\alpha+\frac{2}{(n-1)(n-2)}\Lambda
=\frac{\Omega^2}{n(n-1)}\hat R+O(\Omega^3).
\end{eqnarray}
%
Since we can write $\nabla_a \bar n^a$ as
%
\begin{eqnarray}
\nabla_a \bar n^a=-(n-1)\hat N+\frac{\Omega}{\hat N}\hat
\nabla_a \hat n^a-\frac{\Omega}{\hat N^2}\hat n^a \hat \nabla_a \hat N,
\end{eqnarray}
%
we obtain Eq (\ref{eq:expansion}) in the text. In the same way, we
can derive Eq (\ref{eq:fluc}).


\end{document}